\documentclass[10pt]{article}
\usepackage{spconf,amsmath,graphicx}
\usepackage{cite}
\usepackage{multicol}
\usepackage{caption}
\usepackage{booktabs}
\usepackage{amsfonts}
\usepackage{subcaption}
\usepackage{mwe}
\usepackage{multirow}
\usepackage{titlesec}
\titlespacing*{\section}{0pt}{1.1\baselineskip}{\baselineskip}

\title{Discrete Cosine Transform Based Causal Convolutional Neural Network for Drift Compensation in Chemical Sensors}

\name{Diaa Badawi$^\ast$ \thanks{This work is being supported in part by NSF grants 1739396 (UIC) and 1739451 (ASU).  The authors, Badawi and Cetin, additionally thank NVIDIA for an equipment grant.}\ Agamyrat Agambayev$^\dagger$
	Sule Ozev$^\dagger$ \ A. Enis \c{C}etin$^\ast$}
\address{$^\ast$ Department of Electrical \& Computer Engineering, University of Illinois at Chicago, Chicago, IL \\ $^\dagger$ School of Electrical, Computer and Energy Engineering, Arizona State University, Tempe, AZ}

\begin{document}
	
	\maketitle
	
	\begin{abstract}
		\renewcommand{\baselinestretch}{0.9}
		Sensor drift is a major problem in chemical sensors that requires addressing for reliable and accurate detection of chemical analytes. In this paper, we develop a causal convolutional neural network (CNN) with a Discrete Cosine Transform (DCT) layer to estimate the drift signal. In the DCT module, we apply soft-thresholding nonlinearity in the transform domain to denoise the data and obtain a sparse representation of the drift signal. The soft-threshold values are learned during training. Our results show that DCT layer-based CNNs are able to produce a slowly varying baseline drift signal. We train the CNN on synthetic data and test it on real chemical sensor data. Our results show that we can have an accurate and smooth drift estimate even when the observed sensor signal is very noisy.
	\end{abstract}
	\begin{keywords}
		chemical sensor drift, chemical sensor, time series analysis, discrete cosine transform, convolutional neural networks
	\end{keywords}
	\vspace{-1em}
	\section{Introduction}
	\vspace{-1em}
	Chemical sensors have been used for detecting and identifying chemical analytes in a wide array of industrial and safety applications\cite{arshak2004review}. %They provide an easily portable and cheap alternative to other chemical compound detection methods such as gas chromatography and mass spectrometry \cite{arshak2004review}.
	Despite the fact that chemical sensor technology provides practical solutions, sensor responses may degrade with time, resulting in inconsistent results.%it comes with its own shortcomings [1],[2].
	%One of the major problems with chemical sensors is that their responses vary over time.
	This phenomenon is known as \emph{sensor drift}. Sensor drift arises because of internal factors such as sensor poisoning and aging, and/or external factors such as humidity and temperature changes \cite{holmberg1996drift}. 
	%The former is known as first-order drift, whereas the latter is known as second-order drift. 
	The chemical sensory system becomes unreliable over time, if the sensor drift signal is not properly estimated.
	%Without addressing the issue of sensor drift, one compromises the reliability of the sensory system. 
	
	There has been extensive work to address the drift problem using signal processing and machine learning. One common approach is to extract features carefully from the time series sensor measurements and optimize a machine learning algorithm to recognize patterns from these features \cite{vergara2012chemical,badawi2019detecting}. In \cite{vergara2012chemical}, discriminative features %such as the difference between the maximum and the minimum sensor readouts at different filtering levels
	are extracted from the responses of the sensors and then fed into support vector machines to identify the gas analyte identity. Others have employed different methods to counteract the sensor drift for the task of analyte identification, such as domain adaptation \cite{liu2013drift, zhang2014domain}, semi-supervised learning \cite{de2012semi}, and deep stacked autoencoders \cite{liu2015gas}. In the aforementioned papers, the features extracted from the sensors become unreliable as the sensor ages. 
	%which justifies the low recognition rates for the sensors after operating for long. 
	On the other hand, other approaches try to estimate the drift directly from the time-series measurements of the sensors using iterative interpolation techniques \cite{huang2009reconstruction}, or by using Independent Component Analysis \cite{di2002counteraction}. After the drift signal is estimated, it can be subtracted from the sensor measurements to determine the actual sensor response.
	
	The wide success of deep convolutional neural networks (CNNs) has been demonstrated by their ability to achieve state-of-the-art performance in many time-series recognition tasks. %, among which are time-series related recognition tasks, such as classification, prediction, and interpolation.
	CNNs have been used in analyte classification problems \cite{peng2018gas, badawi2019detecting}. % Furthermore, CNNs have been increasingly preferred to recurrent neural networks in time-series recognition problems, owing to their high architectural flexibility and their relative ease to train.
	One particular variant of convolutional neural networks, the temporal convolutional neural network (TCNN), has been shown to outperform recurrent neural networks on many benchmark data sets \cite{bai2018empirical,mokatren2019deep}. This motivates us to employ a TCNN-based framework to address the problem of drift correction in chemical sensor data.
	Given the fact that sensor drift is a slowly varying signal, 
	we propose utilizing the \emph{Discrete Cosine Transform} (DCT) to extract a baseline drift signal from the observed sensor signal. 
	%We smooth the processed data by applying a soft-threshold to the DCT domain features. 
	%The threshold value is determined during backpropagation based network training. 
	In particular, we propose a new type of layer to be integrated into the conventional TCNN, which we call DCT-based sub-network. In the DCT-based sub-network, we compute the DCT over sliding short-time windows of the temporal feature maps. We then apply the soft-thresholding nonlinearity in the transform domain and compute the inverse DCT to transform the features back to time domain. 
	%As pointed above, t
	In the transform domain, we use soft-thresholding that is parametrized by threshold variables, which are learned during training by the standard backpropagation algorithm. Our results show that the the nonlinear denoising and smoothing in transform domain using the DCT-based structure removes the high-frequency features, i.e., induces smoothness in the features and, subsequently, the constructed output baseline drift signal. Any significant deviation from the sensor baseline signal indicates the existence of chemical vapors.
	
	The organization of this paper is as follows: First we explain the drift problem, TCNN-framework and the DCT-based structure for drift estimation in Section 2.  In Section 3, we present experimental results. Finally, in Section 4 we conclude our work.
	\vspace{-1em}
	\section{Causal DCT-based Real-Time Sensor Drift Estimation Framework}
	\vspace{-0.8em}
	%Drifting phenomenon in ChemFETs has been studied extensively in the literature \cite{ChCh08, ChChCh10}. 
	The drift response of ChemFET sensors is modeled as a linear combination of exponential decays \cite{ChCh08, ChChCh10, BoRoBe83, GoGeCo11, JaMuBr08}. In our ChemFET
	sensors %(without surface coating), 
	we can approximate the drift waveform using the  following  model \cite{karabacak2016making, chengmo2020}:
	\vspace{-.05in}
	\begin{equation} 
	\label{eq:exponential_id}
	d(t) = (R_0+ \epsilon_R) + R_{f} \  \text{exp}\Big(\frac{-t}{\tau_{f}+\epsilon_f}\Big) +
	R_{s} \ \text{exp}\Big(\frac{-t}{\tau_{s}+\epsilon_s}\Big)
	\end{equation}
	\noindent where $R(0) = R_0 + \epsilon_R + R_f + R_s$ is the sensor conversion factor at the initial time after deployment and reset, $R_{f}$, $R_{s}$ are fast and slow drift coefficients, $\tau_{f}$, $\tau_{s}$ are fast and slow drift time constants, and $\epsilon_R$, $\epsilon_f$, $\epsilon_s$ are the corresponding error terms for each of the model variables. %\footnote{The fast and slow drift time constants are set based on hardware measurements on  ChemFETs which have no specific surface coating (hence do not target a specific molecule) \cite{karabacak2016making}. However, ChemFET drift behavior is similar with or without the surface coating as the same binding mechanism applies.}. 
	%The fast and slow drift coefficients each have a nominal value of 0.2 with a standard deviation of 10\% to account for process variations. The fast and slow drift time constants are set to 10s and 500s, respectively. Each time constant is also assigned a 10\% standard deviation to account for process variations\footnote{While the drift coefficients and time constants are based on hardware measurements, the process variation model is based on experience with CMOS parameteric variations (e.g. threshold voltage for large feature-size devices). }.
	It is not possible to know the values of the time constants and other parameters in Eq. \ref{eq:exponential_id} but 
	this model enables us to generate a random population of sensors with individual sensitivities and individual time constants that can differ significantly. 
	%Thus, each sensor's usable} time will vary depending on its own assigned parameters.  
	
	Let a sensor time measurement be $y(t)$ at time $t$. Let $d(t)$ be the drift signal at time $t$. Let $p(t)$ be the desired sensor response without the drift component (drift-corrected response in absence of noise). The observed sensor measurement is $y(t) = d(t)+p(t)+v(t)$, where $v(t)$ is the noise signal.
	Given $y(t)$, we are interested in estimating the drift signal $d(t)$ so that we can estimate $p(t)$.
	
\begin{figure}[t]
		\centering
		\includegraphics[width=0.8\linewidth, height=0.3\linewidth]{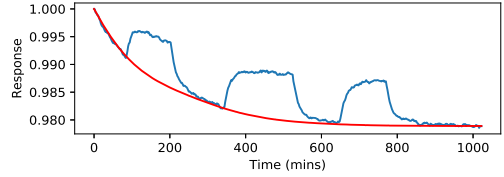}
		\caption{Overall sensor measurement signal (blue) and the underlying drift signal (red)}
		\label{fig:example}
	\end{figure}

	When a sensor is exposed to a gas vapour, 
	it starts absorbing the vapour and 
	its response changes accordingly. Once the sensor is no longer exposed to the gas vapour, it starts exuding the vapour it had absorbed earlier. Absorbing and exuding the vapour depends on the sensor material, the analyte type and the environment. 
	%According to \cite{carmel2003feature}, 
	The ideal sensor response in the absence of drift can be approximated analytically as follows \cite{carmel2003feature},:
	\begin{equation}\label{Eq:analytic_response}
	\resizebox{0.95\linewidth}{!}{
		$p(t) = \begin{cases}
		0  &t \leq T_s \\
		\beta \tau \ \text{tan}^{-1} \big( \frac{t-T_s}{\tau} \big)    &T_s \leq t \leq T_s+\Delta T \\
		\beta \tau \big[  \text{tan}^{-1} \big( \frac{t-T_s}{\tau} \big) - \text{tan}^{-1} \big( \frac{t-T_s-\Delta T}{\tau} \big) \big]   & t \geq T_s+\Delta T
		\end{cases}$}
	\end{equation}
	where $T_s$ is the starting time of the exposure, and $\Delta T$ is the exposure duration. When the sensors are not exposed to any chemical analytes, the response will be simply $y(t)=d(t)+v(t)$. Therefore, we can consider estimating the drift as a baseline (background) estimation problem. An example is shown in Fig \ref{fig:example} in which the sensor is exposed to Volatile Organic Compounds (VOC) three times (the blue curve). Since the drift waveform is decaying, it is not possible to set a threshold without estimating the drift waveform to detect the gas. 
	
	In \cite{huang2009reconstruction}, the drift estimation is posed as a missing data interpolation problem and the Papoulis-Gerchberg (PG) method is used. The PG method is an iterative method involving sequential forward and inverse Fourier Transform computations over blocks of sensor data. It requires some time-domain information about the drift signal and it is assumed that the sensor is not exposed to gas vapor initially.
	%prior knowledge as to when the sensor is exposed to the gas vapour. Henceforth, 
	To interpolate the ``missing'' drift portions where $p(t) \neq 0$, the authors also assumed a low-pass bandwidth (BW) for the drift signal in the Fourier domain \cite{huang2009reconstruction}. 
	In contrast, we do not assume any prior bandwidth information. The neural network  automatically imposes the sparsity constraint on the drift signal in the transform domain during training by learning DCT domain soft thresholds. Furthermore, we do not use future samples during the DCT computation to estimate the drift signal $d(t)$.
	We use only the past and current samples of $y(t)$ to estimate the drift signal $d(t)$. As a result, we compute a real-time estimate of $d(t)$\footnote{We could have used DFT instead of DCT but we preferred the DCT over DFT because the DCT is a real-valued transform.}.
	\\
	%Nevertheless, our estimates for $d(t)$ and $p(t)$ need to satisfy certain constraints. The first constraint is that $d(t)$ be low-frequency, i.e. $\int_{t=0}^\infty d(t) e^{-j \omega t} dt =0$ for $|\omega| >\text{BW}$, where $\text{BW}$ is the signal bandwidth. The second constraint is that $p(t) \geq 0$, i.e. the sensor response when the drift is eliminated is non-negative. Finally, $d(t)$ and $p(t)$ should be of bounded variation.
	%\vspace{-1em}
	{\bf 2.1 TCNN with a DCT layer}
	
	We developed a Temporal Convolutional Neural Network (TCNN) with a DCT layer to extract %both $p(t)$ and 
	the drift signal
	$d[n]$ from the observed sensor signal $y[n]$. In TCNN, the convolutional layers implement causal convolution. Therefore, in this approach, given a time series $y[t]$ for $t = \{1, 2, \ ... \  n\}$, one can determine a drift estimate $\hat{d}[n]$ and, subsequently, $\hat{p} [n] = y[n] - \hat{d}[n]$ \footnote{We will use the discrete-time notation from now on}.
	TCNN implements the so-called dilated convolution in their convolutional layers, which can be expressed as $y_r[n] := \sum_{k=0}^{K-1} h[k] x[n-rk],$
	%\begin{equation}
	%y_r[n] := \sum_{k=0}^{K-1} h[k] x[n-r %k],
	%\end{equation}
	where $h$ is the 1-d temporal filter of length $K$, and $r$ is the dilation rate. 
	
	TCNN is made up of successive convolutional blocks, for which the dilation rate of the convolution is increased exponentially as we go up. Furthermore, residual connections are also used between successive blocks of convolutional layers. %In practice, the dilation rate is increase exponentially as we go to a higher block, which greatly reduces the number of layers needed to reach a large dilation rate with respect to the length of the input sequence.
	%The use of dilated convolution along with residual connections equips the network with the ability to learn long-term and short term patterns for its task. 
	This is particularly useful for our task given that the drift signal is a low frequency signal. On the other hand an exposure to a gas analyte vapor will result in a sudden increase in the recorded sensor signal value $y[n]$. 
	
	Our TCN design goes as follows: We first apply an ordinary convolutional layer with filters of length 5. Afterwards, we apply convolutional blocks with dilated convolution and residual connections. Each of the convolutional blocks comprises two dilated convolutional layers with a spatial dropout layer in between. The second layer in each block has residual connections with the output of the previous block. The dilation rate increases by 2 from $r=2$ in the first dilated convolutional block to $r=2^7=128$ in the last dilated convolutional block. The number of feature maps is 64 throughout all the layers. The spatial dropout rate is set to 0.2. Afterwards, we feed the 64 feature maps of the last layer of TCNN module into the DCT layer, in which we carry out the transform domain processing as explained in Sec. \ref{Sec:DCT}. The DCT window size is set to 64. Finally, we apply a $1\times1$ convolution without any nonlinearity to estimate the drift signal $d[n]$.
	The DCT domain processing is described in Subsection 2.2 in detail.
	
	The whole framework is trained to minimize the following cost function:
	\begin{equation}\label{Eq:cost}
	\begin{split}
	\mathcal{J}:= &\sum_{n=0}^{N-1} \big(d[n]-\hat{d}[n]\big)^2 + %\sum_{n=0}^{N-1} \big( p(n) - \hat{p}(n)\big)^2 + \\ 
	\lambda 
	\sum_{n=1}^{N-1} |\hat{d}[n]-\hat{d}[n-1]|, %\lambda_2 \sum_{n=1}^{N-1} |\hat{p}(n)-\hat{p}(n-1)|    
	\end{split}
	\end{equation}
	where the first term in the cost criterion is the reconstruction square errors. The second term is the Total Variation (TV) regularization term.\\
	%for $\hat{d}[n]$.\\
	%\vspace{-1em}
	{\bf 2.2 DCT-based Layers}
	\label{Sec:DCT}
	
	Discrete cosine transform has been widely used in image compression and speech coding. In this work, we propose incorporating  DCT in the TCNN framework to construct smooth drift estimates from the observed signal $y[n]$. 
	In particular, let $f_i[n]$ be the i-th feature arising from the TCNN at time instance $n$.  The DCT coefficients  %$F_{n,i} (k), \ k=0,1,.., N-1$ 
	over a sliding windows of size $N$ are given by:
	\vspace{-1em}
	\begin{equation}\label{Eq:DCT_short}
	F_{n,i} (k
	) := \sum_{l=0}^{N-1} f^i[n-l] \text{cos} \Big(\frac{\pi k}{N} (N-l+\frac{1}{2})\Big),
	\end{equation}
	where $k\in \{0, 1, ... , N-1\}$ is the DCT index, and ${F}_{n,i} $ is the transform of the feature segment $f_i[t]  \ \text{for} \ t \in \{n-N+1, ..., n\}$. %Given a new sample at time $n+1$, we will have a corresponding feature $f_i[n+1]$ that will be realized by the TCNN.
	%We then will apply DCT according to Eq. \ref{Eq:DCT_short}, resulting in a new set of $N$-point spectral features that will be processed subsequently.
	\begin{figure}
		\centering
		\includegraphics[width=\linewidth]{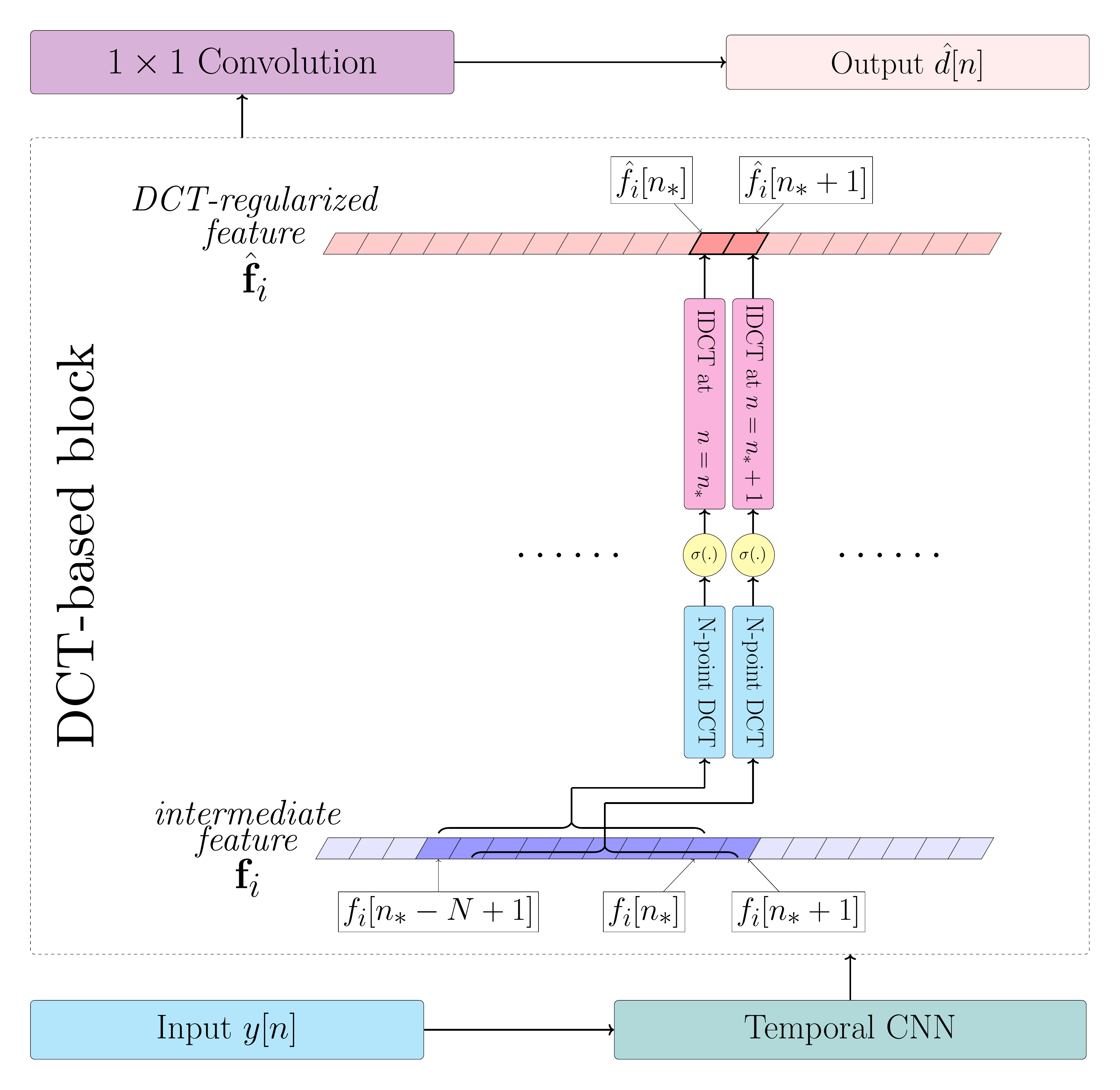}
		\caption{Block diagram of the DCT-based NN structure. The non-linearity $\sigma(.)$ represents to the soft-thresholding function.}
		\label{fig:dct_struct}
	\end{figure}
	After the DCT computation we apply sign-preserving soft-thresholding nonlinearity in the transform domain. %Otherwise, any asymmetric nonlinearity would result in dramatic deformation of the features signal when transformed back into time domain, rather than simply smoothing it.
	For this purpose, we use the soft thresholding function 
	defined for a scalar input $x$ as $\texttt{SoftTh}(x) := \text{sgn}(x) \text{max}(0,|x|-b)$.
	%\begin{equation}\label{Eq:soft_thresholding}
	%   \texttt{SoftTh}(x) = \begin{cases}
	%      x-b, \quad &x\geq b \\
	%    0, \quad   &|x| \leq b \\
	%      x+b, \quad   &x \leq -b
	%   \end{cases},
	%\end{equation}
	The threshold $b\geq 0$ is analogous to the additive bias term in regular time-domain convolutional layers. The parameter $b$ is learned during training by standard backpropagation. The soft-thresholding function $\texttt{SoftTh}(.)$ is applied element-wise to the coefficients ${F}_{n,i} (k)$. Each feature map $f_i$ will have an associated thresholding parameter $b_i$. This proposed soft-thresholding also induces sparsity in the spectral representation. After thresholding we transform back to the time (feature) domain using the inverse discrete cosine transform (IDCT) and realize a smoothed feature point $\hat{f}_{i}[n]$. Notice that, for each feature map, we only need to evaluate IDCT for one single point at time $n$, that is the smoothed version of the original feature value at time $n$. The DCT based TCNN structure is shown in Fig. \ref{fig:dct_struct}.
	\begin{table*}[!htbp]
		%\begin{table}[]
		\centering
		%\resizebox{\columnwidth}{!}{%
		\begin{tabular}{c|c|c|c|c|c|c|c|c|c|c|c|c|c}
			\multirow{2}{*}{Metric}& \multicolumn{12}{c}{Bandwidth used in the PG Algorithm $(\times \frac{2 \pi}{4096})$}& \multirow{2}{*}{TCNN-DCT}  \\
			
			&5&6&7&8&9&10&11&12&13&14&15&16 \\
			\midrule
			MSE (avg.) &1.47&0.91&0.79&1.23&1.34&1.16&0.99&0.86&1.03&1.25&1.54&1.70&\textbf{0.05} \\
			MSE (median) &0.28&0.10&0.23&0.22&0.20&0.16&0.16&0.17&0.27&0.33&0.37&0.55&\textbf{0.02}\\
			\hline
			Cos sim (avg.)&0.81&0.87&0.89&0.77&0.74&0.83&0.87&0.89&0.85&0.82&0.79&0.72&\textbf{0.97} \\
			Cos sim (median)&0.94&0.95&0.92&0.87&0.89&0.92&0.93&0.93&0.88&0.85&0.80&0.74&\textbf{0.99} \\
			
		\end{tabular}
		%}
		\caption{Comparison of  TCNN-DCT framework vs the PG algorithm over the 5,000 examples in the validation data set.  We implemented the PG algorithm with different bandwidth values. 
			%The frequency resolution is $\frac{2 \pi}{4096}$. 
			We report the results for two metrics: MSE (mean square error between the estimate drift and the true drift) and Cos sim (cosine similarity between the estimated drift and the true drift).}
		\label{tab:val}
	\end{table*}
	%After the DCT layer we have additional regular network layers in the TCNN. 

	%in order to combine the DCT-regularized features into a one time series estimate of the drift of length 512.

	%\subsection{Deep Framework for Signal}
	%Going back to our original objective: estimating $d(t)$ and $p(t)$ from the observed signal $y(t)$, we propose using a two sub-network framework for this purpose. The first part is a TCN tasked to estimate the drift-free signal $p(t)$ from $y(t)$. The result is denoted $\hat{p}(t)$. We then concatenate $y(t)$ and $\hat{p}(t)$ into a 2-channel time series and feed it to the second sub-network, which is supervised to output a drift estimate $\hat{d}(t)$. The second sub-network comprises TCN blocks followed by the DCT-based structure explained in the previous subsection. A visual demonstration of the overall system is shown in Fig \ref{fig:overall_sys}.
	%\begin{figure}[!tbp]
	%    \centering
	%    \includegraphics[width=\linewidth]{overall_system}
	%    \caption{Caption}
	%    \label{fig:overall_sys}
	%\end{figure}
	%\footnote{It is worth mentioning that we are trying to estimate both the drift and the pulse signals earlier. However, our results did not suggest any improvement in estimating the drift signal over simply supervising the neural network to estimate only the drift}. %and $\hat{p}(n)$. 
	\vspace{-1em}
	
	%The synthetic data points are then normalized to match the statistics of the real data. Finally, we re-normalize both the real data and the synthetic data using the global mean and the global standard deviation to yield data points with reasonable ranges. It is worth mentioning that we do not normalize each data point by its own statistics because that would require knowing the entire time series in advance, which violates the causality assumption of our framework.  
	\vspace{-0.8em}
	\section{Experimental Results}
	\vspace{-1em}
	In our experiments we used Electronic-nose (E-nose) measurements from the data collected by the JPL \cite{zhou2006nonlinear} using 32 different carbon-polymer sensors 
	%developed by JPL 
	for air quality monitoring. 
	We only considered 
	recordings
	corresponding to single-gas (methanol) exposure experiments as in \cite{huang2009reconstruction}. 
	Some recording examples are shown in Fig. 3.
	We also collected our own data using the commercially available MQ-137  ammonia sensor which has detection range of 5-500 ppm. Three typical sensor recordings are shown at the bottom three rows of Fig. 3.
	We also generated synthetic data for training and validating our network model.
	The real E-nose data was then used as our test set. All the data are resized to a length of 512.

	%we refer to this data set as \emph{Data Set I}. 
	%The \emph{Data set II} was collected by the Jet Propulsion Laboratory (JPL) \cite{zhou2006nonlinear} using 32 different carbon-polymer sensors 
	%developed by JPL 
	%for air quality monitoring. 
	%We only considered recordings  corresponding to single-gas (methanol) exposure experiments as in \cite{huang2009reconstruction}. 
	%This is because our objective is to test the efficacy of the our drift compensation system, rather than identifying different mixtures of analytes.
	\begin{table}[]
		\centering
		\begin{tabular}{c|c|c|c|c}
			Sensor& \multicolumn{2}{c}{PG Algorithm}& \multicolumn{2}{c}{TCNN-DCT} \\
			& MSE&Cos. sim.& MSE& Cos. sim.\\
			\hline
			Sensor 1&0.07&0.97&0.04&0.98\\
			Sensor 2& 0.03&0.97&0.01&0.99\\
			Sensor 3&0.18&0.99&0.04&0.99\\
			Sensor 4&0.20&0.94&0.12&0.94\\
			Sensor 5&0.04&0.97&0.01&0.99\\
			Sensor 6&0.05&0.89&0.09&0.92\\
		\end{tabular}
		\caption{Comparison of TCNN-DCT framework vs. the PG algorithm over the real data shown in Fig. \ref{Fig:dataset2_res} (We report the results of only 6 sensors due to the lack of space). 
			%We manually estimated the drift for each example. 
			% The bandwidth used in the PG algorithm is also selected manually for each example. %We report the results for two metrics: MSE and Cos. sim. as in Table \ref{tab:val}
		}
		\label{tab:test}
	\end{table}
	We synthesized 10,000 recordings that resemble idealized sensor responses for different analyte exposure profiles with slowly varying drift to train the CNN. We used Eq. \ref{Eq:analytic_response} to synthesize sensor responses and we randomized the starting 
	%points
	instances of gas vapors, the exposure period, the parameters $\beta$ and $\tau$, and the number of exposure sessions in order to create as diverse a data set as possible. We synthesized the drift signal simply by sampling from a Gaussian process with covariance  $\mathbb{E}(x(t_1)-\mu)(x(t_2)-\mu) = \text{exp}\big(-\big({\frac{t_1-t_2}{\alpha}}\big)^2\big)$ 
	%for time instances at time steps $s$ and $t$, 
	where $\alpha$ is chosen to be sufficiently large so that the realizations of the drift signal are slowly varying with time. We normalized both the synthetic and real data using global mean and standard deviation.
	%statistics in order to have time series within reasonable ranges.
	%\vspace{-1em}
	%\subsection{Results}
	We trained our network using the synthetic training data for 80 epochs\footnote{We used mini batches of size 32. We used Adam Optimizer with a learning rate equal to $10^{-3}$, $\beta_1=0.9$, and $\beta_2=0.99$.}.  We set the TV parameter $\lambda$ in Eq. \ref{Eq:cost} to 0.1.
	\begin{figure}[!tbp]
		\centering
		\includegraphics[width=0.9\linewidth]{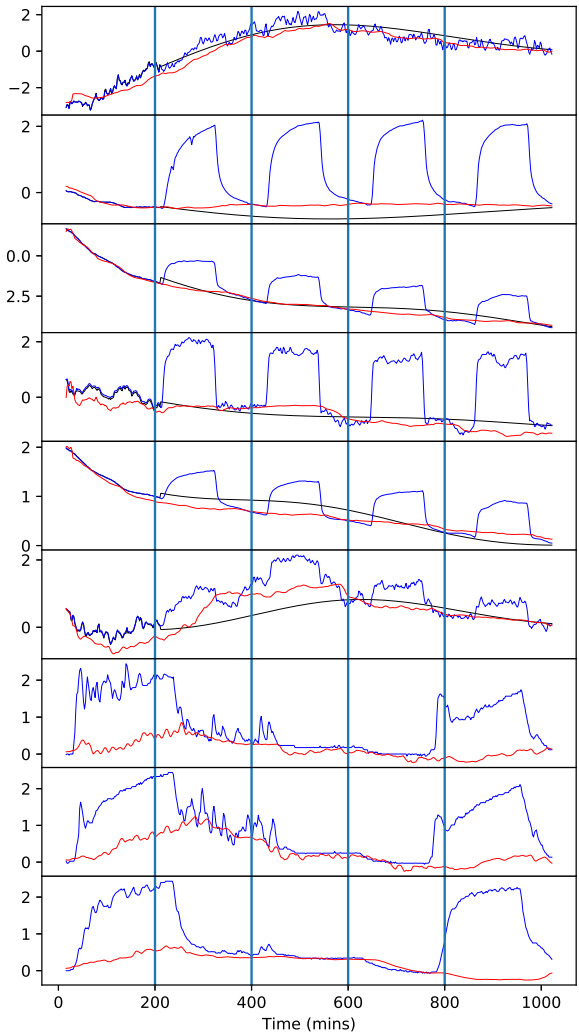}
		%to be added to the text
		\caption{9 real sensor data (in blue), the estimated drift using TCN-DCT framework (in red), and the estimated drift using PG algorithm (in dashed black). The first 6 sensors are from Data Set II, while the last three are from our Data Set I.  The first (top) and the sixth examples are "non-informative"  (damaged) sensors.}
		%Refer to the text for more information}
		%\caption{Real sensor responses (in blue) and the estimated drift (in red). The first six figure correspond to measurements from Data set II, while the last three correspond to measurements from our data set (Data set I). In data set II, there are 4 sessions of vapour exposure as can be seen clearly from Example 1-4. There are two exposure sessions in the measurements from data set II. The segments in dashed black correspond to the interpolated "missing" part of the drift using Papoulis-Gerchberg algorithm. Examples 5 and 6 correspond to measurements of two non-informative sensors. The time scale is in minutes.}
		\label{Fig:dataset2_res}
	\end{figure}
	We implemented the Papoulis-Gerchberg (PG) algorithm over the test data set as in \cite{huang2009reconstruction} to compare it with our approach. We assumed that there is no gas vapor exposure  to the sensor initially and at the end of the recording. %exposure to gas vapor and the last portion of the signal after the last exposure as our available drift data. 
	We extrapolated the remaining part of the drift signal using the PG algorithm. Drift signal estimation results are shown in Fig \ref{Fig:dataset2_res}. 
	%Based on the results in Fig. \ref{Fig:dataset2_res}, 
	The proposed TCNN-DCT framework is superior to the PG algorithm
	in terms of cosine similarity and the MSE as shown in Tables 1 and 2, and Fig. 3. 
	In Table \ref{tab:val}, we have a range of bandwidth values for the PG algorithm to get the best possible result and in Table \ref{tab:test} we manually optimized the bandwidth for the PG algorithm. 
	In Fig. 3, the 1st (top) and the 6th sensors are called "non-informative" (damaged) sensors in \cite{huang2009reconstruction}. The 1st sensor does not produce any response to the gas exposure. In this case, the estimated drift signal is just a smooth (denoised) version of the sensor signal. 
	%
	%in which the the pulses have very distorted shapes, and the drift part is very noise. Nevertheless, the drift is well separated from the overall signal in the second half of the the 6th example. 
	In the last three examples, there is almost no drift, and our drift estimate follows the baseline accurately. 
	%We also report numerical comparative results over the synthetic data in Table \ref{tab:val} and the real data in Table \ref{tab:test}. 
	
	Another advantage of our method compared to the PG method is that we do not require any  prior knowledge about the sensor signal. In the PG method one has to know a gas vapor free segment of the response to estimate the drift.
	
	%In the case of validation data, it is obvious that one needs to find an optimal bandwidth value per example, and global bandwidth choices lead to poor results. This hampers the applicability of the PG algorithm in automatic drift separation. 
	
	\vspace{-1em}
	%{\bf 4.Conclusion}
	\section{Conclusion}
	\vspace{-1em}
	
	We proposed a novel deep DCT structure. The DCT layer is integrated into deep CNNs in order to causally estimate the slowly-varying drift signal in chemical sensors from the sensor response to gas vapor exposure.  In the DCT layer, we perform soft thresholding before transforming back to time domain. 
	The DCT helps regularize the intermediate features generated by the early layers of the TCNN and this leads to an accurate baseline drift signal.
	%We trained our framework over synthetic data that resembles real sensory data with different drift profiles and different gas exposure scenarios, and tested it over real sensor measurements. 
	%Our results show that the drift construction is smooth, but also very accurate. 
	Our network outperforms the PG type sensor drift estimation algorithms without requiring any prior knowledge about the drift signal.

	\bibliographystyle{IEEEtran}
	\vspace{-1em}
	\newpage
	\newpage
	\bibliography{references}
	
\end{document}